\documentclass{aastex}
\input epsf
\received{17 March 2000}
\accepted{XX May 2000}

\slugcomment{Submitted to Ap.J. Letters}

\shortauthors{Crowther et al.}
\shorttitle{Evolutionary status of Sand~2}

\begin{document}

\title{Far-UV FUSE spectroscopy of the O\,{\sc vi} resonance doublet
in Sand~2 (WO)}

\author{Paul~A.~Crowther\altaffilmark{1},
        A.~W.~Fullerton\altaffilmark{2,3},
        D.~J.~Hillier\altaffilmark{4},
        K.~Brownsberger\altaffilmark{5},
        L.~Dessart\altaffilmark{1,6}, 
        A.~J.~Willis\altaffilmark{1},
        O.~De~Marco\altaffilmark{1}, 
        M.~J.~Barlow\altaffilmark{1},
        J.~B.~Hutchings\altaffilmark{7},
        D.~L.~Massa\altaffilmark{8},
        D.~C.~Morton\altaffilmark{7},
        G.~Sonneborn\altaffilmark{9}}

\altaffiltext{1}{Department of Physics \& Astronomy, University 
College London, Gower Street, London WC1E 6BT, UK}
\altaffiltext{2}{Dept. of Physics \& Astronomy, 
                 University of Victoria,
                 P.O. Box 3055, 
                 Victoria, BC, V8W 3P6, 
                 Canada.}
\altaffiltext{3}{Department of Physics \& Astronomy, 
Johns Hopkins University, 
3400 North Charles St., Baltimore, MD 21218}
\altaffiltext{4}{
Department of Physics \& Astronomy, University of Pittsburgh, 
3941 O'Hara Street, PA 15260}
\altaffiltext{5}{
Center for Astrophysics \& Space Astronomy, Campus Box 389, Boulder, CO 80309}
\altaffiltext{6}{
Present Address: D\'{e}partement de Physique, 
Universit\'{e} Laval and Observatoire du Mont M\'{e}gantic, 
Qu\'{e}bec, QC, Canada, G1K 7P4}
\altaffiltext{7}{
D.A.O., Herzberg Institute of Astrophysics,
National Research Council of Canada, 5071 W.~Saanich Road, Victoria BC, 
Canada, V8X 4M6}
\altaffiltext{8}{Raytheon ITSS, NASA's Goddard Space Flight Center,
                 Code 631, Greenbelt, MD 20771}
\altaffiltext{9}{
Laboratory for Astronomy and Solar Physics, NASA/GSFC, Code 681, Greenbelt,  
MD 20771}

\begin{abstract}
We present {\it Far-Ultraviolet Spectroscopic Explorer (FUSE)} spectroscopy
of Sand~2, a LMC WO-type Wolf-Rayet star, revealing the
O\,{\sc vi} resonance P~Cygni doublet at 1032--38\AA.
These data are combined with {\it HST}/FOS ultraviolet and Mt Stromlo 2.3m
optical spectroscopy, and analysed using a spherical,
non-LTE, line-blanketed code. Our study reveals exceptional stellar 
parameters: $T_{\ast}\sim$150,000K, $v_{\infty}$=4100 km\,s$^{-1}$,
log\,($L/L_{\odot})$=5.3, and
$\dot{M}=1\times 10^{-5} M_{\odot}$yr$^{-1}$,
if we adopt a volume filling factor of 10\%. Elemental abundances of
C/He$\sim0.7\pm0.2$ and O/He$\sim0.15_{-0.05}^{+0.10}$ by number 
qualitatively support previous recombination line studies. 
We confirm 
that Sand~2 is more chemically enriched in carbon than LMC WC stars,
and is expected to undergo a supernova explosion within the 
next 5$\times 10^{4}$yr.
\end{abstract}


\section{Introduction}

Wolf-Rayet (WR) stars provide keys to our understanding of massive stellar
evolution, nucleosynthesis processes and chemical enrichment of the
ISM. Of these, the oxygen-sequence (WO) introduced by 
Barlow \& Hummer (1982) is by far the rarest. Their high excitation 
oxygen emission lines are widely interpreted as revealing the late core
helium-burning or possibly carbon-burning stage 
(Smith \& Maeder 1991), of importance for constraining the 
controversial $^{12}$C($\alpha, \gamma)^{16}$O reaction rate.

In spite of searches in all Local Group galaxies, only six massive WO
stars are known to date, namely Sand~1 (Sk\,188) in the SMC, Sand~2
(BAT99--123) in the LMC, Sand 4 (WR\,102),  Sand 5 (WR\,142)
and MS4 (WR\,30a) in our Galaxy, and  DR1 in IC1613.  
Since O\,{\sc vi} 3811--34\AA\ is a primary 
WO classification diagnostic, with an equivalent width of up to 1700\AA\ 
(Kingsburgh, Barlow \& Storey (1995, hereafter KBS), 
observations of O\,{\sc vi} 1032--38\AA\ are keenly sought. 
However, its location in the far-UV has ruled out such observations to date. 
This situation has changed following the successful launch of the 
{\it Far-Ultraviolet Spectroscopic Explorer (FUSE}, Moos et al. 2000),
which permits routine high dispersion far-UV spectroscopy of 
massive stars in the Magellanic Clouds. 

In this Letter, we analyse {\it FUSE} spectroscopy of Sand~2 (Sanduleak 1971), alias
Sk$-$68$^{\circ}$ 145 = Brey~93 = BAT99-123 (Breysacher, Azzopardi \& 
Testor 1999), together with {\it Hubble Space Telescope (HST)} and ground-based datasets. 

\section{Observations}

Previously unpublished far-UV, UV and optical/near-IR spectroscopy of 
Sand~2 have been obtained  with {\it FUSE}, {\it HST} and the 
Mt Stromlo and Siding Spring Observatory (MSSSO) 2.3m, respectively.

\subsection{Far-UV spectroscopy}

Sand~2 was observed  by {\it FUSE} as part of the 
Early Release Observation programme X018 on 1999 Oct 31. A 8134 sec
exposure of Sand~2 with the $30^{''}\times30^{''}$
(LWRS) aperture provided data at $R\sim$12,000 with the two Lithium 
Fluoride (LiF) channels, covering $\lambda\lambda$979--1187\AA, obtained in
time-tag (TTAG) mode. At this epoch, the two Silicon Carbide (SiC) 
channels, covering $\lambda\lambda$905--1104\AA, were badly aligned so 
that SiC data were of poor quality.

Sand~2 data were processed through the pipeline data reduction,
CALFUSE  (version 1.6.8), and are shown in Fig.~1. The
pipeline extracted the 1D spectrum, removed the background, and corrected 
for grating wobble and detector drifts. No corrections for astigmatism 
or flat fielding have been applied. From Fig.~1, the far-UV continuum 
flux of Sand~2 is low
($\approx 5 \times 10^{-14}$ erg\,s$^{-1}$\,cm$^{-2}$\,\AA$^{-1}$), with
two principal stellar features, O\,{\sc vi} $\lambda$1032-38
and C\,{\sc iv} $\lambda$1168. The latter is blended 
with C\,{\sc iii} $\lambda$1175, while 
C\,{\sc iv} $\lambda$1108  and O\,{\sc vi} $\lambda$1125 are 
present, though weak. The {\it FUSE} spectrum is affected by
a multitude of interstellar absorption features, principally 
H\,{\sc i} and H$_{2}$, plus airglow emission features due to
Ly$\beta$, [O\,{\sc i}] and [N\,{\sc i}].

\subsection{Near-UV spectroscopy}

Sand~2 was observed with the {\it HST} Faint Object Spectrograph 
(FOS) instrument during 1996 March (PI: D.J. Hillier, Program ID 5460). 
Exposures totalling 5980, 3780 and 1090 sec were obtained with
the G130H, G190H and G270H 
gratings, respectively. Many important 
spectral features are identified in the {\it HST}/FOS dataset (see also KBS),
principally C\,{\sc iv} $\lambda$1550, C\,{\sc iii} $\lambda$2297,
He\,{\sc ii} $\lambda$1640, O\,{\sc iv} $\lambda$1340, O\,{\sc v} 
$\lambda$1371, $\lambda$2781--87. Following Prinja, Barlow \& Howarth
(1990) we find $v_{\infty}$=4100 km\,s$^{-1}$ from C\,{\sc iv} $\lambda$1548--51
(KBS derived 4500 km\,s$^{-1}$ from {\it IUE} spectroscopy).

\subsection{Optical spectroscopy}

We have used the Double Beam Spectrograph (DBS) at the 2.3m MSSSO
telescope  to observe Sand~2 on 1997 Dec 24--27. Use of a
dichroic and the 300B, 600B and 316R gratings permitted simultaneous
spectroscopy covering 3620--6085\AA\ and 6410--8770\AA\ for 1200\,sec, 
plus 3240--4480\AA\  and 8640--11010\AA\  for 2500\,sec. A 2$''$ slit 
and 1752$\times$532 pixel SiTE CCD's provided a
2 pixel spectral resolution of $\sim$5\AA. A standard data reduction was 
carried out, including absolute flux calibration, using wide slit 
spectrophotometry of HD\,60753 (B3\,IV) and $\mu$ Col (O9\,V), plus
atmospheric correction using HR\,2221 (B8\,V). Convolving our dataset 
with Johnson broad-band filter profiles reveals V=15.1 and B-V=+0.54 mag. 
Since these are contaminated by emission lines, we have 
convolved our data with Smith $ubvr$ narrow-band filters to reveal 
$v$=16.1, $u-b=-$0.19, $b-v=-0.07$ and $v-r=-0.14$ mag.

Our optical dataset confirms the spectral morphology previously presented
and discussed by KBS, with exceptionally strong 
O\,{\sc iv} $\lambda$3400, O\,{\sc iv} $\lambda$3818, C\,{\sc iv}+He\,{\sc ii} 
$\lambda$4660 and C\,{\sc iv} $\lambda$5801. Depending on the WO 
selection criteria, its spectral 
type is WO3 (Crowther, De Marco \& Barlow 1998) or WO4 (KBS).

\section{Quantitative analysis of Sand~2}

Wolf-Rayet winds are so dense that non-LTE effects,
spherical geometry and an expanding atmosphere are 
minimum assumptions.

\subsection{Analysis technique}

We employ the code of Hillier \& Miller (1998), {\sc cmfgen}, which 
iteratively solves the  transfer equation in the co-moving 
frame subject to statistical and radiative  equilibrium in an expanding, 
spherically symmetric and steady-state atmosphere. These models account
for clumping, via a volume filling factor, $f$,  and line blanketing,
both of which have a significant effect on the physical 
properties of WC stars (e.g. Hillier \& Miller 1999). Through the 
use of `super-levels', extremely 
complex atoms can be included. For the present application, a total 
of 3,552 levels (combined into 796 super-levels), 60 depth points and 
63,622 spectral lines of 
He\,{\sc i-ii}, C\,{\sc ii-iv}, O\,{\sc ii-vi}, Ne\,{\sc ii-iv},
Si\,{\sc iv}, S\,{\sc iv-vi}, Ar\,{\sc iii-v}, and 
Fe\,{\sc iv-viii} are considered simultaneously (see 
Dessart et al. 2000 for the source of atomic data used).

We adopt a form for the 
velocity law (Eqn~8 from Hillier \& Miller 1999) such that two exponents 
are considered ($\beta_{1}$=1, $\beta_{2}$=50, 
$v_{\rm ext}$=2900 km\,s$^{-1}$, $v_{\rm turb}$=100 km\,s$^{-1}$), 
with the result that acceleration is modest at  small radii, 
but continues to large distance (0.9$v_{\infty}$ is reached at 100$R_{\ast}$). 

Our spectroscopic analysis derives $\dot{M}/\sqrt{f}$, 
rather than $\dot{M}$ and $f$ 
individually, since line blending is severe in Sand~2. 
A series of models were calculated in which stellar 
parameters, $T_{\ast}$, log\,$(L/L_{\odot})$,
$\dot{M}/\sqrt{f}$, C/He and O/He, were adjusted until  the 
observed line strengths and spectroscopic fluxes were
reproduced. 
We adopt 0.4$Z_{\odot}$ abundances for Ne, Si, S, Ar 
and Fe. The distance to the LMC was assumed to be 51.2\,kpc 
(Panagia et al. 1991).  

The wind ionization balance is ideally deduced using isolated spectral 
lines from adjacent ionization stages  of carbon (C\,{\sc iii} 
$\lambda$2297, C\,{\sc iv} $\lambda$5801-12) or oxygen (O\,{\sc iv}
$\lambda$3404-14, O\,{\sc v} $\lambda$3144, O\,{\sc vi} $\lambda$5290). 
In practice this was difficult to  achieve for Sand~2 because of the 
severe line blending. 

\subsection{Stellar parameters}
Our initial parameter study of Sand~2 revealed a fairly similar 
spectral appearance spanning 120kK$\le T_{\ast}\le$170kK, with 
log\,$(L/L_{\odot})$=5.28 and log\,($\dot{M}/M_{\odot}$yr$^{-1})=-$4.94 fixed. 
The principal differences between these models are that (i) the 
observed strength of the 
O\,{\sc vi} 3811--34\AA\ doublet, plus lines in the red such as 
C\,{\sc iv} 7700\AA\ favour a high $T_{\ast}$; 
(ii) the weakness of the O\,{\sc vi} 1032--38\AA\ doublet, as revealed by 
{\it FUSE}, favours a lower $T_{\ast}$. Since other parameters, in particular
abundances, are largely unaffected by these discrepancies, we 
shall adopt $T_{\ast}$=150kK (i.e. $R_{\ast}$=0.65$R_{\odot}$). 
Note that higher luminosity models do produce significant effects, 
such as a dramatic weakening of C\,{\sc iv} 5801--12\AA\ emission.

Fig.~2 compares our synthetic spectrum with the observed 
far-UV, UV and optical spectroscopy of Sand~2. Our model is reddened 
by $E(B-V)$=0.08 mag due to our Galaxy, obtained from the 
reddening map of Burstein \& Heiles (1982). An additional LMC
component of 0.11 mag was required, such that $M_{v}=-3.0$\,mag.
In the absence of a far-UV extinction law, standard UV laws 
(Seaton 1979; Howarth 1983) are {\it extrapolated} for $\lambda\le$1200\AA\
with (variable) influence on the fit quality to {\it FUSE} data. 

Overall the observed spectrum of Sand~2 is very well reproduced 
by the model, with most He\,{\sc ii}, C\,{\sc iii-iv} and O\,{\sc iv-vi} 
lines matched in strength and shape, except for O\,{\sc vi} 
$\lambda\lambda$3811--34 (model too weak) and 
O\,{\sc vi} $\lambda\lambda$1032--1038 (model too strong) as 
discussed above. In particular, the flat-topped nature 
of C\,{\sc iii} $\lambda$2297 is well matched. It is clear that 
although WO stars have little or no C\,{\sc iii} $\lambda$5696
(a classification diagnostic), other C\,{\sc iii} lines are indeed
present (Hillier 1989). Many spectral features in Sand~2 are due to 
blends because of the very 
broad spectral lines and the fact that recombination lines 
of O\,{\sc vi} and C\,{\sc iv} overlap with He\,{\sc ii} lines.
For example, the spectral feature at $\lambda\lambda$4650--4686 has principal
contributors He\,{\sc ii} $\lambda$4686, C\,{\sc iii} $\lambda\lambda$4647-50, 
C\,{\sc iv} $\lambda$4658 and C\,{\sc iv} $\lambda$4685, while minor 
contributors include C\,{\sc iv} 
$\lambda$4646, $\lambda$4689 and O\,{\sc vi} $\lambda$4678.

He\,{\sc ii} $\lambda$5412/C\,{\sc iv} 
$\lambda$5471  provides an excellent diagnostic of C/He (e.g. 
Hillier \& Miller 1998) for spectroscopic studies of WC stars.
However, the large line widths of
WO stars, and the fact that C\,{\sc iv} $\lambda$5412 (14--8) contributes
to He\,{\sc ii} $\lambda$5412 (7--4) hinders
the use of these lines (see inset box in Fig.~2). Instead, we are 
able to derive C/He$\sim0.7\pm$0.2 by number from He\,{\sc ii} (6--4)
$\lambda$6560/C\,{\sc iv} $\lambda$7700, although C\,{\sc iv} (12--8)
contributes to the former.
Other C\,{\sc iv} and He\,{\sc ii} recombination lines show excellent 
agreement, except that C\,{\sc iv} $\lambda$1107 is predicted to 
be stronger than {\it FUSE} observations reveal.

In contrast to WC stars, numerous oxygen recombination lines are
present in the UV and optical spectra of WO stars (e.g. 
O\,{\sc vi} $\lambda$5290, $\lambda$2070). From their strength 
relative to He and C recombination lines, we estimate 
O/He$\sim0.15_{-0.05}^{+0.10}$ by number. The weakness of 
lower ionization oxygen features, such as O\,{\sc iv} $\lambda$1400 and 
O\,{\sc v} $\lambda$3150 also argue against higher O/He ratios, 
although O\,{\sc v} $\lambda$5590 favours O/He$\sim$0.25.
Poor fits to O\,{\sc vi} $\lambda\lambda$1032--38 and $\lambda\lambda$3811--34 
are discussed above, while O\,{\sc v} $\lambda\lambda$2781--87 and 
O\,{\sc iv} $\lambda\lambda$3063--71 are systematically too weak for all models.

We present the
predicted temperature structure of our Sand~2 model in
Fig.~3. Although Sand~2 is an extremely hot star, its 
very high content of efficient C and O coolants (see 
Hillier 1989) directly results in a very cool
($<$10kK) outer wind ($r/R_{\ast} >$ 100).
Consequently, the ionization structure of metal species is predicted to 
be very stratified, such that O$^{6+}$ is the dominant ionization stage 
for $r/R_{\ast} <$5, yet O$^{3+}$ is dominant for $r/R_{\ast} >$25.

The high stellar temperature of Sand~2 implies a high bolometric 
correction ($-$5.7 mag) and consequently hard ionizing spectrum, 
such that 50\% of the emergent photons have energies greater
than 13.6eV (912\AA\ Lyman edge), and 30\% greater than 24.6eV (504\AA\
He\,{\sc i} edge).  The ionizing fluxes in the H\,{\sc i}, 
He\,{\sc i}, and He\,{\sc ii} continua are 
10$^{49.1}$ s$^{-1}$, 10$^{48.8}$ s$^{-1}$ and 10$^{40.3}$ 
s$^{-1}$, respectively. WO stars are known to produce strong 
nebular He\,{\sc ii} $\lambda$4686 emission in associated 
H\,{\sc ii} regions (e.g. Kingsburgh \& Barlow 1995), although the 
predicted number of He\,{\sc ii} continuum ionizing photons in our
Sand~2 model is much lower than those inferred from other WO stars.

\section{Discussion}

Abundances derived here qualitatively support the results from 
KBS, who used recombination line theory to derive C/He=0.5 and O/He=0.1
for Sand~2. WO stars are well suited to recombination studies for 
carbon,  although oxygen is somewhat more problematic for recombination 
line studies, since the ionization structure is more complex 
(see Fig.~3), and few lines have available coefficients.

Gr\"{a}fener, Hamann \& Koesterke (1999) have carried out a detailed 
non-LTE spectroscopic analysis of Sand~2 (see Gr\"{a}fener et al. 1998). 
Overall, we confirm their $\dot{M}/\sqrt{f}$ determination, but
derive a higher luminosity (by 0.2\,dex) and temperature 
(they derived $T_{\ast}$=101kK), attributable to the incorporation of 
line blanketing. More significantly, we obtain systematically lower 
metal abundances (they estimated C/He=1.3 and 
O/He=1.2), such that  the oxygen mass fraction for Sand~2 is only 
16\%, versus 50\% according to Gr\"{a}fener et al. (1999). We attribute 
this major revision to improved spectroscopic and atomic datasets 
(Gr\"{a}fener et al. used simple C and O model atoms plus low 
S/N optical data). Our higher luminosity and clumpy wind conspire 
to revise the 
wind performance ratio, $\dot{M} v_{\infty} / (L/c)$, from 56 to 12.

Fig.~4 compares the luminosities and (C+O)/He ratios for Sand~2 
with six LMC WC4 stars, updated from Dessart (1999). He 
improved upon similar work by Gr\"{a}fener et al. (1998) using 
line blanketed, clumped models, revealing a greater range of 
carbon abundances, 0.1$\le$C/He$\le$0.3 (due to improved
spectroscopy), systematically higher luminosities (because of 
blanketing and improved reddenings), and lower mass-loss rates (due 
to clumping). The carbon enrichment of Sand~2 is substantially higher 
than the WC4 stars,
although O/He does not differ so greatly from the WC4 sample, for which
O/He$\le$0.08. This supports the suggestion by Crowther (1999) 
that unusually high oxygen enrichment may not be a pre-requisite for a 
WO classification. 

We have superimposed (non-rotating) evolutionary tracks from
Meynet et al. (1994) at 0.4$Z_{\odot}$ for $M_{\rm initial}$=60, 
85 and 120$M_{\odot}$ on Fig.~4. These evolutionary models predict 
C/O$\sim$2 when C/He$\sim$0.7, in conflict with our determination
of C/O$\sim$4. Better agreement is expected for evolutionary models 
in which rotation is accounted for, since these predict higher 
C/O ratios during the WC/WO phase (Maeder \& Meynet 2000). 
From interior models, $M_{\rm initial}\ge 60 M_{\odot}$  for Sand~2,
with a corresponding age of $\sim$3--4.3\,Myr, such that a supernova 
is expected within the next 0.1--5$\times 10^{4}$ years.
The stellar luminosity implies a current mass of 10$M_{\odot}$ 
(Schaerer \& Maeder 1992), such that the {\it mean} 
post-main sequence mass-loss rate is 
2.5--5$\times 10^{-5}$ $M_{\odot}$\,yr$^{-1}$.

Quantitative analysis of other WO stars suffer from either (i) 
high interstellar 
reddening (WR\,102, WR\,142), (ii) complications because of binarity 
(Sand~1), or (iii) large distances (DR1). Nevertheless, we expect 
similar C and O enrichment to that derived here for Sand~2 (KBS).

\acknowledgments

This work is based on data obtained: (i) for the Guaranteed Time Team by the
NASA-CNES-CSA FUSE mission, operated by the Johns Hopkins University,
(ii) with the NASA/ESA Hubble Space Telescope obtained from the 
data archive at the STScI, which is operated by AURA under the NASA 
contract NAS5-26555; (iii) with the 2.3m Mount Stromlo and Siding Spring
Observatory. Financial support is acknowledged from the Royal Society (PAC),
NASA contract NAS5-32985 (U.S. FUSE participants), STScI grant 
GO-04550.01-92A, NASA grant NAG5--8211 (both JDH), PPARC (OD) 
and the UCL Perren Fund (LD).

\clearpage

\figcaption[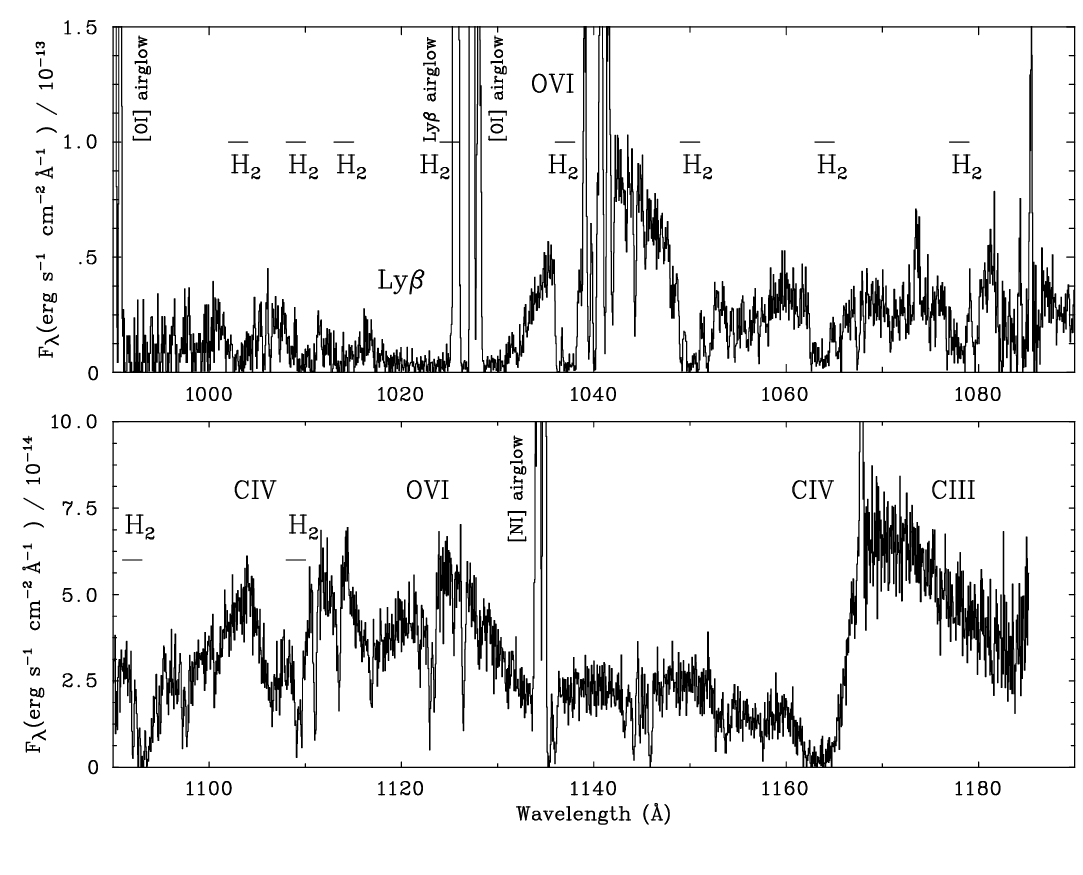]{Observed {\it FUSE} (CALFUSE 1.6.8) spectroscopy 
of Sand~2, rebinned to 15 km\,s$^{-1}$, revealing the P~Cygni 
O\,{\sc vi} resonance doublet at 1032--38\AA. The far-UV spectrum
is affected by interstellar atomic and molecular hydrogen lines, 
plus airglow from [O\,{\sc i}], [N\,{\sc i}] and Ly$\beta$.\label{fig1}}

\figcaption[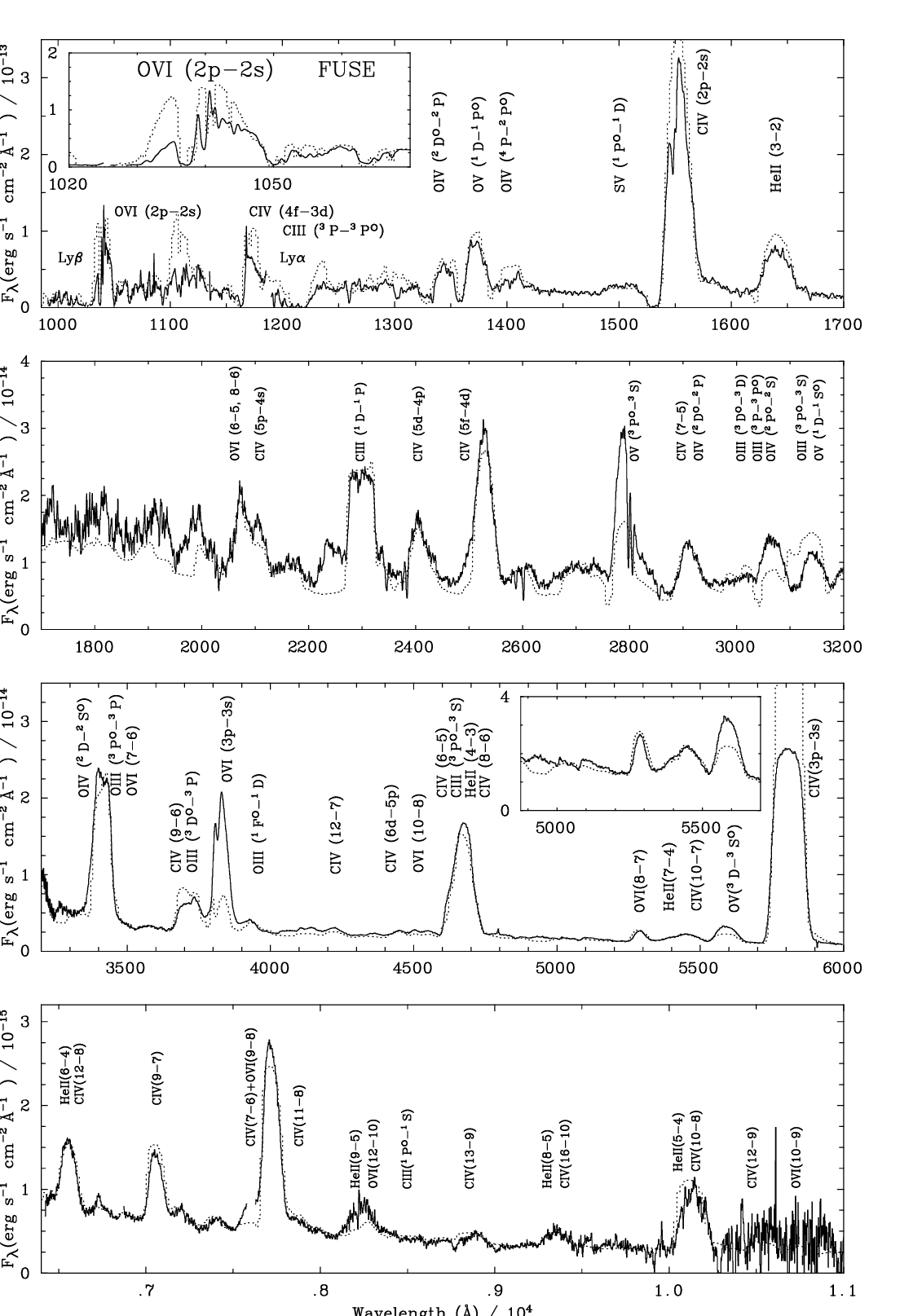]{Comparison between far-UV ({\it FUSE}), 
UV ({\it HST}/FOS), optical and near-IR (MSSSO 2.3m) spectroscopy of
Sand~2, shown as a solid line, together with our synthetic spectrum
(dotted line), reddened by 
$E(B-V)$=0.08 (Gal) and 0.11 (LMC). Correction for interstellar 
H\,{\sc i} and H$_{2}$ in the 
far-UV has been applied following J.E. Herald, D.J. Hillier 
\& R.E. Schulte-Ladbeck (in preparation), for which we adopt
$T_{\rm ISM}$=100K, $v_{\rm turb}$=10 km\,s$^{-1}$. 
We use $\log(n$(H\,{\sc i})/cm$^{2}$)=20.6 (Gal) plus 21.3 (LMC) 
(Shull \& van Steenberg 1985; Koornneef 1982), plus
$\log(n$(H$_{2}$)/cm$^{2}$)=20.\label{fig2}}

\figcaption[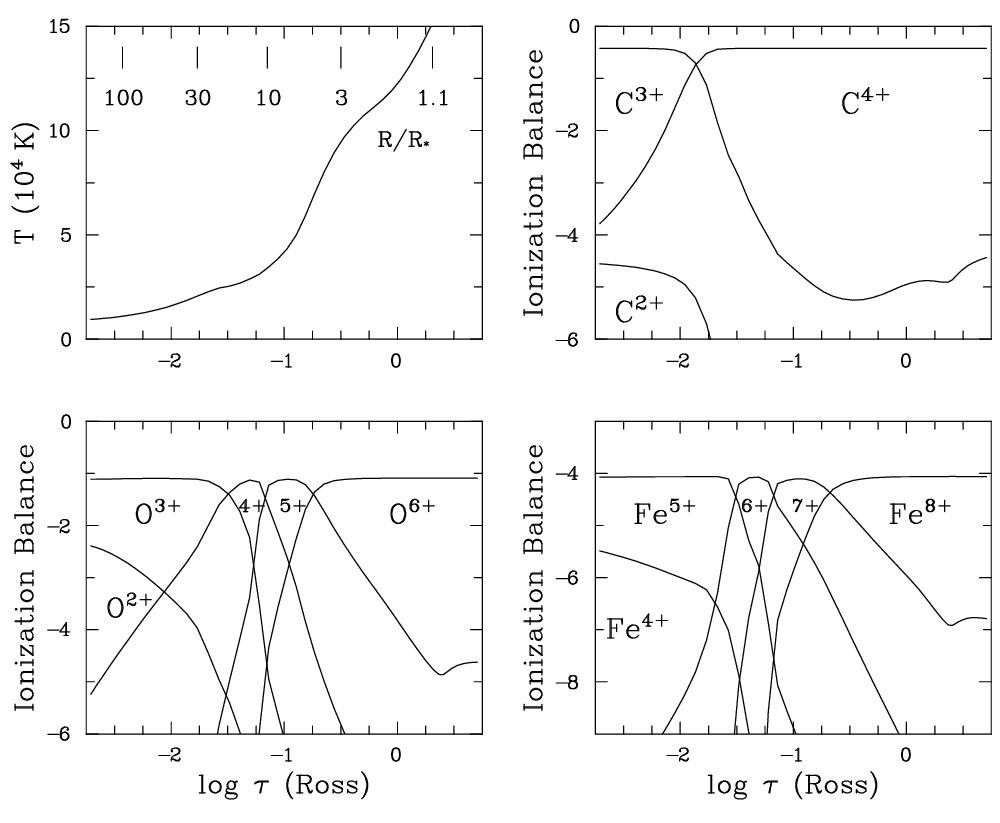]{Theoretical temperature distribution for our Sand~2 model
versus  Rosseland  optical depth, indicating selected stellar radii,
$R_{\ast}$, plus the ionization balance of C, O and Fe.\label{fig3}}

\figcaption[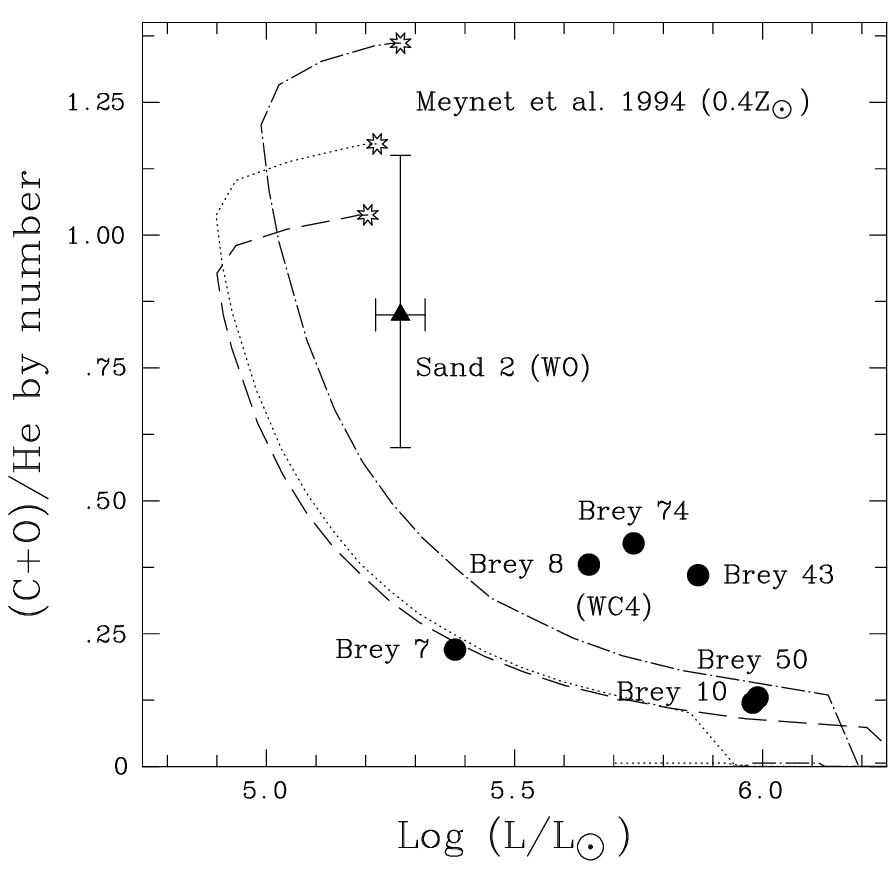]{Comparison between the luminosity (in $L_{\odot}$) and surface 
abundances ((C+O)/He, by number)
of Sand~2 (triangle) with six LMC WC4 stars (circles, Dessart 1999),
plus evolutionary predictions of Meynet et al. (1994) for $M_{\rm initial}$=60 
(dotted), 85 (dot-dashed) and 120 $M_{\odot}$ (dashed), including
locations of SN explosions (stars).\label{fig4}}

\clearpage

\begin{figure}
\epsfxsize=15cm \epsfbox[0 205 520 625]{FIG1.eps}
\end{figure}

\begin{figure}
\epsfxsize=15cm \epsfbox[30 30 573 805]{FIG2.eps}
\end{figure}

\begin{figure}
\epsfxsize=15cm \epsfbox[40 210 520 605]{FIG3.eps}
\end{figure}

\begin{figure}
\epsfxsize=15cm \epsfbox[50 185 480 610]{FIG4.eps}
\end{figure}

\clearpage

\begin{thebibliography}{}
\bibitem[Barlow and Hummer(1982)]{bar82} Barlow, M.J.,\&  Hummer, D.G., 1982, 
in IAU Symp. 99, Wolf-Rayet Stars, Observations, Physics, Evolution, eds.
de Loore, C.W.H., \& Willis, A.J. (Dordrecht: Reidel), 387\
\bibitem[Breysacher et al.(1999)]{brey99} Breysacher, J., Azzopardi, M., \& 
Testor, G., 1999, A\&AS 137, 117
\bibitem[Burstein and Heiles(1982)]{bh82} Burstein, D., \& Heiles, C., 1982, AJ 87, 1165
\bibitem[Crowther(1999)]{cro99} Crowther, P.A., 1999, in IAU Symp. 193, 
Wolf-Rayet Phenomena in Massive Stars and Starburst Galaxies, eds.
van der Hucht K.A., Koenigsberger G., \& Eenens P.R.J., (San Francisco: ASP), 116
\bibitem[Crowther et al.(1998)]{cro98}Crowther,~P.A.,~De~Marco,~O.,~\&~Barlow,~M.J.,~1998,~\mnras,~296,~367
\bibitem[Dessart(1999)]{luc99}Dessart, L., 1999, PhD thesis, University of London
\bibitem[Dessart et al.(2000)]{luc2000} Dessart, L., Crowther, P.A., Hillier,
D.J., Willis, A.J., Morris, P.W., \& van der Hucht, K.A., 2000, MNRAS in press (astro-ph/0001228)
\bibitem[Gr\"{a}fener et al.(1998)]{goetz98} Gr\"{a}fener, G., Hamann, W-.R., 
Hillier, D.J., \& Koesterke, L., 1998, 
\aap, 329, 190
\bibitem[Gr\"{a}fener et al.(1999)]{goetz99a} Gr\"{a}fener, G., Hamann, W-.R., 
\& Koesterke, L., 1999, in IAU Symp. 193, Wolf-Rayet Phenomena in Massive Stars 
and Starburst Galaxies, eds. van der Hucht, K.A., Koenigsberger, G., 
\& Eenens, P.R.J., (San Francisco: ASP), 240
\bibitem[Hillier(1989)]{hil89} Hillier, D.J., 1989, \apj, 347, 392
\bibitem[Hillier and Miller(1998)]{hm98} Hillier, D.J., \& Miller, D.L., 1998, \apj, 496, 407
\bibitem[Hillier and Miller(1999)]{hm99} Hillier, D.J., \& Miller, D.L., 1999, \apj, 519, 354
\bibitem[Howarth(1983)]{how83} Howarth, I.D., 1983, \mnras, 203, 301
\bibitem[Kingsburgh and Barlow(1995)]{robin95b} Kingsburgh, R.L., \& Barlow, M.J., 1995, \aap, 295, 171
\bibitem[Kingsburgh et al.(1995)]{robin95a} Kingsburgh, R.L., Barlow, M.J., \& Storey, P.J., 1995, \aap, 295, 75 (KBS)
\bibitem[Koesterke and Hamann(1995)]{kh95} Koesterke, L., \& Hamann, W-.R., 1995, \aap, 299, 503
\bibitem[Koornneef(1982)]{koo82} Koornneef, J., 1982, \aap, 107, 247
\bibitem[Maeder and Meynet (2000)]{andre00} Maeder, A., \& Meynet, G., 2000, ARA\&A, 38, in press
\bibitem[Meynet et al.(1994)]{georges94} Meynet, G., Maeder, A., Schaller, G.,
Schaerer, D., \& Charbonnel, C., 1994, \aaps,  103, 97
\bibitem[Moos et al.(2000)]{moos2000} Moos, H.W., et al. 2000, ApJ, this issue.
\bibitem[Panagia et al.(1991)]{pan91} Panagia, N., Gilmozzi, R., Macchetto, F., 
Adorf, H.-M., \& Kirshner, R.P., 1991, \apj, 380, L23
\bibitem[Prinja Barlow \& Howarth (1990)]{pbh90} Prinja, R., Barlow, M.J., \& Howarth, I.D., 1990, ApJ 361, 607
\bibitem[Sanduleak(1971)]{sand71} Sanduleak, N., 1971, \apj, 164, L71
\bibitem[Schaerer and Maeder(1992)]{sm92} Schaerer, D.,  \& Maeder, A., 1992, \aap, 263, 129
\bibitem[Seaton(1979)]{mike79} Seaton, M.J., 1979, \mnras, 187, 73
\bibitem[Shull and van Steenberg(1985)]{ss85} Shull, J.M., \& van Steenberg, M.E., 1985, ApJ 294, 599
\bibitem[Smith and Maeder(1991)]{sm91} Smith, L.F., \& Maeder, A., 1991, \aap, 241, 77
\end{thebibliography}
\end{document}